\documentclass[conference]{IEEEtran}
\IEEEoverridecommandlockouts
\usepackage{cite}
\usepackage[utf8]{inputenc}
\usepackage{amsmath,amssymb,amsfonts}
\usepackage{algorithmic}
\usepackage{graphicx}
\usepackage{tabularx}
\usepackage{textcomp}
\usepackage{xcolor}
\usepackage{multirow} 
\usepackage{array}  
\usepackage{tikz}
\usepackage{float} 
\usepackage{hyperref}
\usepackage{xcolor}
\usepackage{orcidlink}
\hypersetup{
    colorlinks=true,
    linkcolor=blue,
    citecolor=blue,
    filecolor=magenta,      
    urlcolor=cyan,
    pdftitle={},
    pdfpagemode=FullScreen,
}

\usetikzlibrary{shapes.geometric, arrows, positioning}
\def\BibTeX{{\rm B\kern-.05em{\sc i\kern-.025em b}\kern-.08em
    T\kern-.1667em\lower.7ex\hbox{E}\kern-.125emX}}
\begin{document}

\title{Leveraging Cloud-Fog Automation for Autonomous Collision Detection and Classification in Intelligent Unmanned Surface Vehicles}
\author{\IEEEauthorblockN{
    Thien Tran\IEEEauthorrefmark{1},
    Quang Nguyen\IEEEauthorrefmark{2}\IEEEauthorrefmark{3},
    Jonathan Kua\IEEEauthorrefmark{1},
    Minh Tran\IEEEauthorrefmark{3}\IEEEauthorrefmark{6},
    Toan Luu\IEEEauthorrefmark{4}\IEEEauthorrefmark{3},
    Thuong Hoang\IEEEauthorrefmark{1}, and
    Jiong Jin\IEEEauthorrefmark{5}
    } \\
    \IEEEauthorrefmark{1}Deakin University, Australia;
    \IEEEauthorrefmark{2}University of Birmingham, UK;
    \IEEEauthorrefmark{3}RMIT University, Vietnam;\\
    \IEEEauthorrefmark{4}VinUniversity, Vietnam;
    \IEEEauthorrefmark{5}Swinburne University of Technology, Australia;
    \IEEEauthorrefmark{6}University of Tasmania, Australia \\
    {\{peter.tran, jonathan.kua, thuong.hoang\}@deakin.edu.au};
    {mxn498@student.bham.ac.uk};\\
    {minh.tranquang@rmit.edu.vn};
    {toan.ld@vinuni.edu.vn}; 
    {jiongjin@swin.edu.au} \\
}
\maketitle

\begin{abstract}
Industrial Cyber-Physical Systems (ICPS) technologies are foundational in driving maritime autonomy, particularly for Unmanned Surface Vehicles (USVs). However, onboard computational constraints and communication latency significantly restrict real-time data processing, analysis, and predictive modeling, hence limiting the scalability and responsiveness of maritime ICPS. To overcome these challenges, we propose a distributed Cloud-Edge-IoT architecture tailored for maritime ICPS by leveraging design principles from the recently proposed Cloud-Fog Automation paradigm. Our proposed architecture comprises three hierarchical layers: a Cloud Layer for centralized and decentralized data aggregation, advanced analytics, and future model refinement; an Edge Layer that executes localized AI-driven processing and decision-making; and an IoT Layer responsible for low-latency sensor data acquisition. Our experimental results demonstrated improvements in computational efficiency, responsiveness, and scalability. When compared with our conventional approaches, we achieved a classification accuracy of 86\%, with an improved latency performance. By adopting Cloud-Fog Automation, we address the low-latency processing constraints and scalability challenges in maritime ICPS applications. Our work offers a practical, modular, and scalable framework to advance robust autonomy and AI-driven decision-making and autonomy for intelligent USVs in future maritime ICPS.
\end{abstract}


\section{Introduction}\label{sec:sec1}
The adoption of maritime Industrial Cyber-Physical Systems (ICPS) has significantly advanced Unmanned Surface Vehicles (USVs) by enabling intelligent autonomous operations in navigation, environmental monitoring, and industrial applications~\cite{Tabish2024, Kinaci2023, Feng2018}. As ICPS become integral to maritime autonomy, their ability to detect and classify collisions in low-latency remains a critical challenge. Conventional collision detection systems are based on visual sensors (cameras, LiDAR, and radar), often susceptible to environmental disturbances such as fog, low visibility, and water splashes~\cite{Thombre2022}. In addition, alternative sensing methods that rely on proprioceptive data from the Inertial Measurement Unit (IMU) have emerged, offering robustness against environmental disturbances~\cite{Nath2018}. By utilizing derived signals as input data, Artificial Intelligence (AI)-driven mechanisms, such as Convolutional Neural Networks (CNN) classification methods, can reliably identify collisions in low-latency environments without dependence on visual sensing conditions~\cite{Nguyen2024IMU}. However, onboard processing limits computational resources, resulting in high latency, scalability issues, and energy inefficiency in real-world deployment~\cite{Ieracitano2024}.

Although recent studies have explored IMU-based collision classification for USVs, existing methods are mainly based on localized/centralized processing, limiting low-latency performance and scalability. The recently Cloud-Fog Automation paradigm provides a distributed, intelligent, and scalable computing framework that enables real-time decision-making, adaptive control, and resilient automation in ICPS, particularly in complex, latency-sensitive, or resource-constrained environments \cite{jin2025cfa_autonomous, Lyu2024}. While widely adopted in ICPS, most USV-based ICPS architectures currently do not leverage network-centric architectures. This gap motivates us to propose and develop a distributed Cloud-Edge-IoT architecture tailored for low-latency USV collision classification and detection, while efficiently distributing computational loads, decentralization, and power efficiency~\cite{redowan2025trusted, Shi2016}.

\begin{figure*}[!ht]
    \centering
    \includegraphics[width=0.75\textwidth]{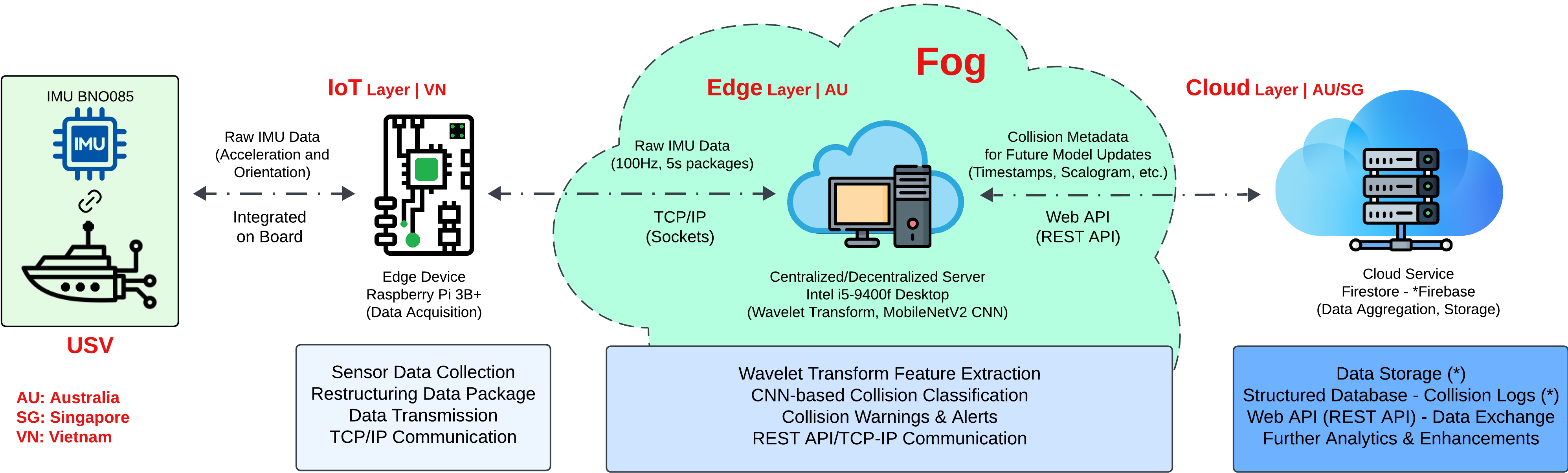}
    \caption{Distributed Cloud-Edge-IoT architecture leveraging Cloud-Fog Automation for USVs' collision detection and classification.}
    \label{fig_CEIoT}
\end{figure*}

In this paper, we make the following key contributions: 

\begin{itemize}
    \item Proposed a hierarchical Cloud-Edge-IoT architecture tailored for maritime ICPS.
    \item Developed a distributed modular computing architecture for low-latency USVs' AI-driven collision classification and detection.
    \item Experimentally validated the classification accuracy, processing time, and latency analysis across hierarchical layers in controlled water environments.
\end{itemize}

The remaining paper is structured as follows: Section~\ref{sec:sec2} presents the proposed system architecture, while Section~\ref{sec:sec3} details the research methodology. Section~\ref{sec:sec4} describes the experimental setup, followed by Section~\ref{sec:sec5}, which presents experimental results and discussions. Finally, Section~\ref{sec:sec6} concludes the paper and outlines future research directions.

\section{System Architecture}\label{sec:sec2}
To address the computational and system performance limitations associated with onboard processing, we propose a distributed Cloud-Edge-IoT computing architecture based on Cloud-Fog Automation to improve power efficiency, scalability, manageability, and complexity constraints. The system is structured into three hierarchical layers: the Cloud Layer (data aggregation for future enhancement), the Edge Layer (AI-driven detection and classification), and the IoT Layer (data acquisition). In this architecture, the fog computing domain encompasses all distributed processing that occurs between the Cloud and IoT layers, enabling adaptive, real-time decision-making close to the data source. This distributed architecture ensures computational load balancing, low-latency decision-making, and scalability for ICPS in maritime applications.

Figure~\ref{fig_CEIoT} illustrates the overall system architecture, highlighting the distinct roles of each layer and the communication flow throughout the network. This hierarchical layered processing ensures that low-latency collision classifications are performed at the Edge layer, minimizing delays while offloading long-term storage functions to the Cloud layer.

The proposed framework integrates proprioceptive sensing IMU, AI-driven classification, and distributed data processing to improve low-latency collision classification for USVs. Each layer is designed to perform specialized tasks, ensuring a seamless and hierarchical computational workflow:

\begin{itemize}
    \item \textbf{Cloud Layer:} Functions as a structured data storage node, maintaining essential records and structured datasets for future analytics, predictive maintenance, retraining of AI models, and human validation reference.
    \item \textbf{Edge Layer:} Conducts feature extraction using Wavelet Transform, applies a CNN model for collision classification, and then provides low-latency warnings.
    \item \textbf{IoT Layer:} Acquires raw proprioceptive IMU data and performs basic data restructuring before transmitting to the Edge Layer for collision classification and detection.
\end{itemize}

\subsection{Cloud Layer: Data Aggregation for Future Enhancement}\label{sec:sec2.A}
The Cloud Layer (hosted on the Cloud Server) serves as a structured data repository for collision event logs, sensor data, and related classification results. Unlike conventional Cloud-based architectures, where AI processing is conducted remotely, this work retains AI inference at the Edge Layer~\cite{Jin2024, Lyu2024} while utilizing the Cloud Layer exclusively for:

\begin{itemize}
    \item Long-term storage of classified collision data.
    \item Dataset aggregation for future AI model refinements and predictive maintenance research.
    \item System-wide analytics for monitoring the performance of the distributed computing architecture.
\end{itemize}

It is noted that the Cloud Layer does not yet perform computations or model retraining, but remains a scalable component for further system enhancements through predictive analytics, and federated learning-driven optimization~\cite{Liu2024, Woo2020}.

\subsection{Edge Layer: AI-driven Classification and Decision-Making}\label{sec:sec2.B}
The Edge Layer (implemented on a centralized/decentralized computing unit) is responsible for data feature extraction and AI-driven collision classification. This layer processes incoming IMU data and classifies collision types using a CNN trained on labeled impact datasets. The Edge Layer facilitates low-latency AI processing by leveraging a distributed and powerful computational infrastructure. Although processing tasks are not localized directly onboard, this approach effectively reduces reliance on cloud connectivity, thus significantly enhancing operational responsiveness and reliability. Key processing steps include:
\begin{itemize}
    \item Wavelet Transform to extract time-frequency domain features from IMU signals.
    \item CNN-based classification model optimized for low-latency inference, distinguishing between different stages of impacts and collision position.
    \item A Decision-making framework that generates collision alerts and action recommendations based on the detection classification results.
\end{itemize}

\subsection{IoT Layer: Data Acquisition and Preprocessing}\label{sec:sec2.C}
The IoT Layer consists of a single-board computer connected to an IMU. The IMU continuously captures linear acceleration and angular velocity data, providing proprioceptive sensing for collision event detection. By offloading complex processing to the Edge Layer, the IoT Layer remains computationally lightweight and energy-efficient, crucial for overall system performance enhancement. To ensure data integrity and low-latency processing, this layer only performs basic data restructuring and data transmission via Wi-Fi.

\section{Methodology}\label{sec:sec3}

\subsection{Data Acquisition and Preprocessing}\label{sec:sec3.A}
The IoT Layer (implemented on a Raspberry Pi 3B+) is responsible for IMU data acquisition and preliminary preprocessing. The system utilizes an Adafruit IMU BNO085, which provides 9-degree-of-freedom (9-DOF) orientation data by capturing 1) Linear acceleration (m/s²), 2) Angular velocity (rad/s), 3) Orientation (Euler angles), etc. To ensure high temporal resolution, the IMU samples data at 100Hz, transmitting the combined data package of a 5-second window with a window offset of 3 seconds to the Edge Layer. The raw data is restructured, packaged, and transmitted from the IoT to the Edge using the TCP/IP protocol. The raw dataset contains noise and interference, mitigated at the IoT layer by the BNO085 IMU's integrated MCU, which performs preprocessing and sensor fusion before data use.

\subsection{Feature Extraction and AI-driven Classification}\label{sec:sec3.B}
The Edge Layer (Intel i5-9400f Desktop) receives proprioceptive IMU data, performs advanced feature extraction using Morlet Wavelet Transform techniques~\cite{Liu2021b}, and classifies collisions utilizing a MobileNetV2-based CNN model.

\subsubsection{Wavelet Transform-Based Feature Extraction}\label{sec:sec3.B.1}
\begin{figure}[!t]
    \centering
    \includegraphics[width=1.\columnwidth]{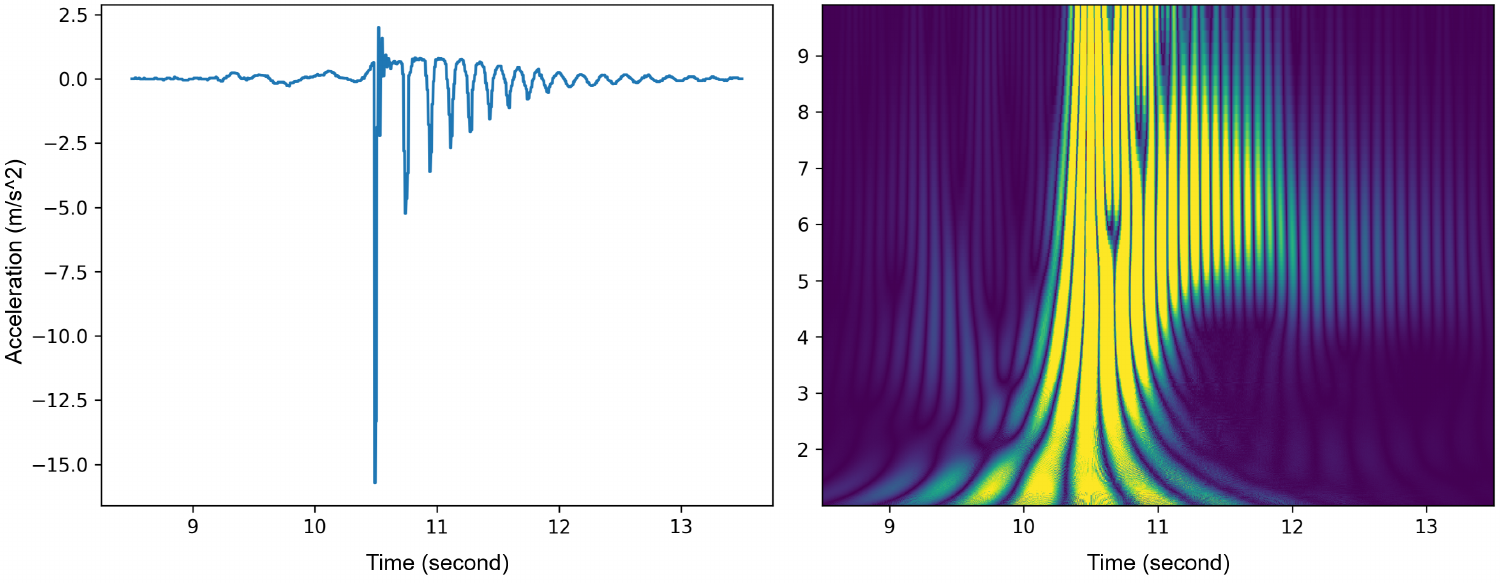}
    \caption{X-axis raw acceleration data (left) and continuous Wavelet Transform scalogram (right) for a ship collision event.}
    \label{fig_raw2cwt}
\end{figure}

To effectively capture both time-domain and frequency-domain patterns, this process employs the Wavelet Transform for feature extraction. This method offers distinct advantages over Fourier-based analysis by providing localized, time-sensitive frequency decomposition. While Fourier analysis provides high-frequency resolution, wavelets balance resolution with precise time localization, making them particularly suitable for transient event detection and making them well-suited for impact identification in USVs~\cite{Liu2021a}. The extracted wavelet coefficients are converted into scalograms, 2D visual representations of energy distribution, and input to the CNN classifier. Figure \ref{fig_raw2cwt} illustrates the application of the continuous Wavelet Transform to ship collision data. The selection of the wavelet function plays a crucial role in determining the resolution of the continuous Wavelet Transform across the time and frequency domains, an aspect that, together with identifying appropriate wavelets for specific signal characteristics.

\subsubsection{Convolutional Neural Network (CNN) Model}\label{sec:sec3.B.2}
\begin{figure}[!t]
    \centering
    \includegraphics[width=0.8\columnwidth]{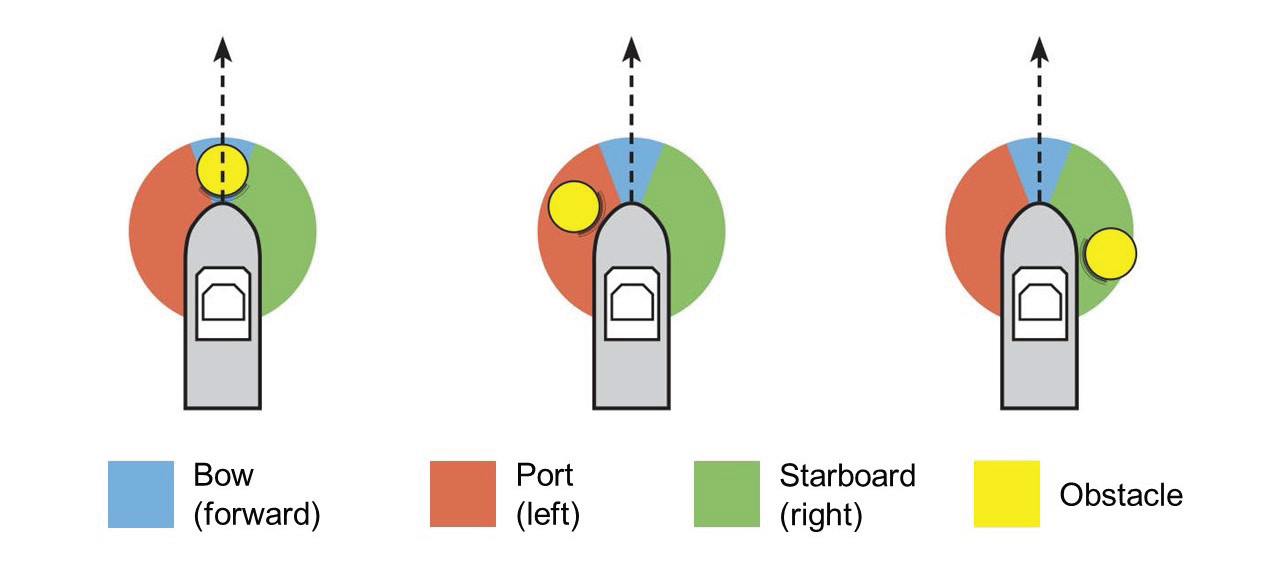}
    \caption{Three main impact types of USV collision classification \cite{Nguyen2024IMU}.}
    \label{fig_imp}
\end{figure}

A custom CNN architecture~\cite{Skulstad2021} is trained on labeled wavelet scalogram, distinguishing between three types of collision and default stage: 1) Bow (Frontal); 2) Port (Left-Side); 3) Starboard (Right-Side); 4) None (Default) as shown in Figure~\ref{fig_imp}. The classification task utilizes a lightweight CNN architecture (MobileNetV2), a convolutional neural network optimized for embedded systems and resource-constrained hardware~\cite{Sandler2018}. The MobileNetV2 CNN model employs inverted residuals with linear bottlenecks, beginning with a standard 3x3~convolutional~layer followed by bottleneck blocks composed of a 1x1~expansion convolution, depthwise convolution, and a 1x1~projection convolution to maintain efficient, low-dimensional representations. It processes six input channels (3~linear~acceleration, 3~angular~acceleration), receives 192x150~RGB~scalograms in a 5-second~duration~(X-axis) and under 10Hz~(Y-axis), with normalized continuous Wavelet Transform coefficient range. These details ensure clarity regarding the temporal length and input channels, enhancing the reproducibility of our research methodology.

\subsection{Low-Latency Communication and System Integration}\label{sec:sec3.C}
The proposed architecture employs TCP/IP and REST API communication protocols over the Internet to facilitate seamless integration~\cite{ladegourdie2022performance}, as illustrated in Figure~\ref{fig_CEIoT}. Specifically, the IoT Layer transmits a restructured IMU data package to the Edge Layer for immediate processing and collision classification. Subsequently, the Edge Layer forwards classified collision metadata and relevant analytical data to the Cloud Layer (Firestore), serving primarily as a repository for data aggregation, long-term analytics, and future scalability.

The data transmission includes timestamped collision events, IMU data, classified collision types, Wavelet Transform outputs, and human-validated reference. This structured communication pipeline ensures robust and efficient integration across all layers for the distributed computing processes, optimizing low-latency responsiveness and analytical capabilities.

\section{Experimental Setup}\label{sec:sec4}

\subsection{Hardware and Software Configurations}\label{sec:sec4.A}
\begin{table}[!b]
\centering
\caption{Experimental setup and configuration details of the distributed Cloud-Edge-IoT computing architecture.}
\label{tab1}
\renewcommand{\arraystretch}{1.3} 
\footnotesize 
\begin{tabularx}{\columnwidth}{|p{0.9cm}|p{1.5cm}|p{1.5cm}|X|}
\hline
\textbf{Layer} & \textbf{Device} & \textbf{Software} & \textbf{Function} \\
\hline
IoT Layer 
& Raspberry \newline Pi 3B+, IMU \newline BNO085
& Raspberry \newline Pi OS, \newline Sockets
& IMU data acquisition \newline (100Hz) and \newline dataset restructuring \\
\hline
Edge Layer & Intel \newline i5-9400f Desktop & TensorFlow, PyWavelets, Python, \newline MobileNetV2 & Feature extraction (Wavelet Transform), CNN classification, low-latency collision decision-making \\
\hline
Cloud Layer & Firestore (Firebase) & Web API, Firebase Database & Structured data storage, \newline aggregation for future analytics and model retraining \\
\hline
\end{tabularx}
\end{table}

The system is implemented using three hierarchical layers, each equipped with distinct hardware and software components. Table~\ref{tab1} below summarizes the system configurations. The IoT Layer collected 100Hz IMU data and transmits processed data to the Edge Layer. The Edge Layer performed Wavelet Transform-based feature extraction and CNN-based collision classification, forwarding classified results to the Cloud Layer for data aggregation and future analytics.

\subsection{Testing Environment and Collision Emulation}\label{sec:sec4.B}
\begin{figure}[!t]
    \centering
    \includegraphics[width=0.59\columnwidth]{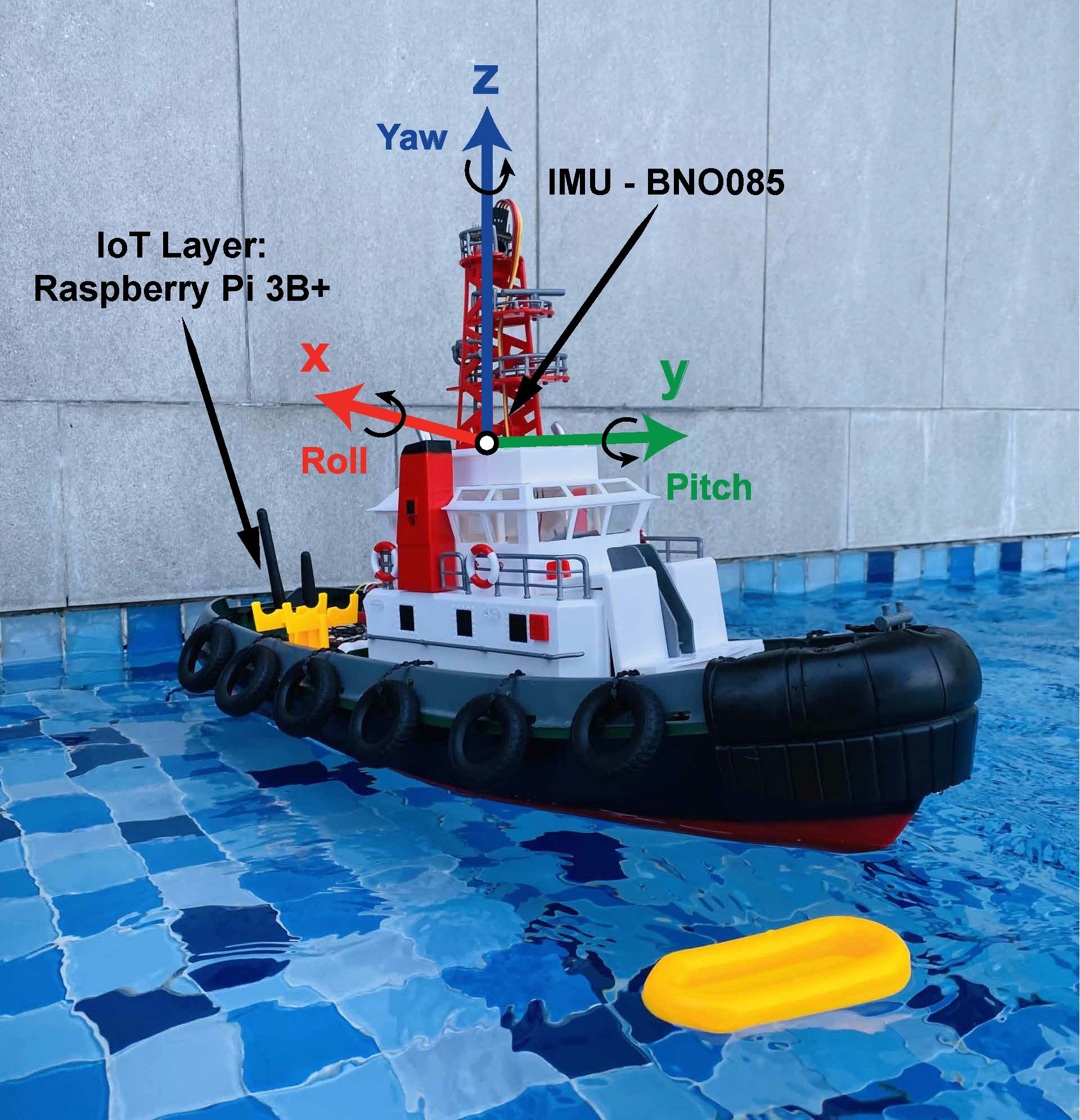}
    \caption{Model-scale USV with integrated onboard system, featuring a BNO085 IMU and Raspberry Pi 3B+.}
    \label{fig_ship}
\end{figure}

To ensure realistic and controlled testing conditions, the experimental setup utilized a model-scale USV deployed in a calm water environment. The testbed is designed to replicate real-world maritime conditions while maintaining repeatability and controlled impact scenarios. Figure~\ref{fig_ship} presents that the USV is fitted with an IMU BNO085 sensor and a Raspberry Pi 3B+ at the end of the shipdeck.

We use a model-scale vessel to represent the USV in our work. The IMU sensor is securely mounted to the hull, ensuring minimal external disturbances affect the measurements. The USV is manually controlled to impact the predefined collision points for data testing collections. To minimize external disturbances, our experiments are conducted in an 8m×4m×2m enclosed water basin to provide a controlled water surface with predictable impact dynamics across multiple trials, and consistency in environmental conditions for repeatability.


The testing framework consisted of three predefined impact zones, replicating real-world collision scenarios. Each scenario is repeated at least 10 times per impact type, ensuring statistical reliability in classification accuracy. The IMU sensor recorded acceleration and angular velocity at the moment of impact, providing raw motion data for analysis. 

\subsection{Evaluation Metrics and Data Collection}\label{sec:sec4.C}
The system performance is evaluated using four primary metrics, as summarized in Table~\ref{tab2}. These metrics are selected to assess the accuracy, responsiveness, and communication efficiency of the proposed system. To ensure reliable evaluation, data is systematically collected during multiple experimental trials conducted under controlled conditions. Each trial involved recording IMU data, classification outputs, and precisely measuring processing time and latency across layers. Metrics are quantified through confusion matrix analysis, system timestamps, and Round-Trip Time (RTT) analysis to provide an overall performance of the proposed system. In this study, the RTT analysis is used to evaluate the full processing time of Edge-to-Cloud data transmission.

\begin{table}[b] 
\centering
\caption{Performance Evaluation Metrics}
\label{tab2}
\small
\renewcommand{\arraystretch}{1.3}
\begin{tabularx}{\columnwidth}{|p{1.8cm}|X|X|}
\hline
\textbf{Metric} & \textbf{Definition} & \textbf{Measurement Method} \\ \hline
Classification \newline Accuracy 
& Correctly classified \newline collision event rate
& Confusion matrix \newline analysis\\ 
\hline
IoT  \newline  Processing Time
& Processing time of \newline IMU data collection and restructuring
& Timestamped \newline processing before \newline transmission to Edge \\
\hline
IoT-to-Edge \newline Latency 
& Transmission time \newline IMU data package \newline from IoT to Edge
& Timestamped transmission \newline events log\\
\hline
Edge Layer \newline Processing Time 
& Processing time of feature extraction and CNN classification 
& Time stamps captured \newline at data arrival and \newline CNN output\\
\hline
Edge-to-Cloud \newline Latency 
& Transmission time \newline from Edge to Cloud \newline for storage
& Measure RTT of \newline structured data \newline transmission\\
\hline
\end{tabularx}
\end{table}

\section{Results and Discussions}\label{sec:sec5}

\subsection{Classification Performance}\label{sec:sec5.A}
\begin{figure}[!ht]
    \centering
    \includegraphics[width=0.47\columnwidth]{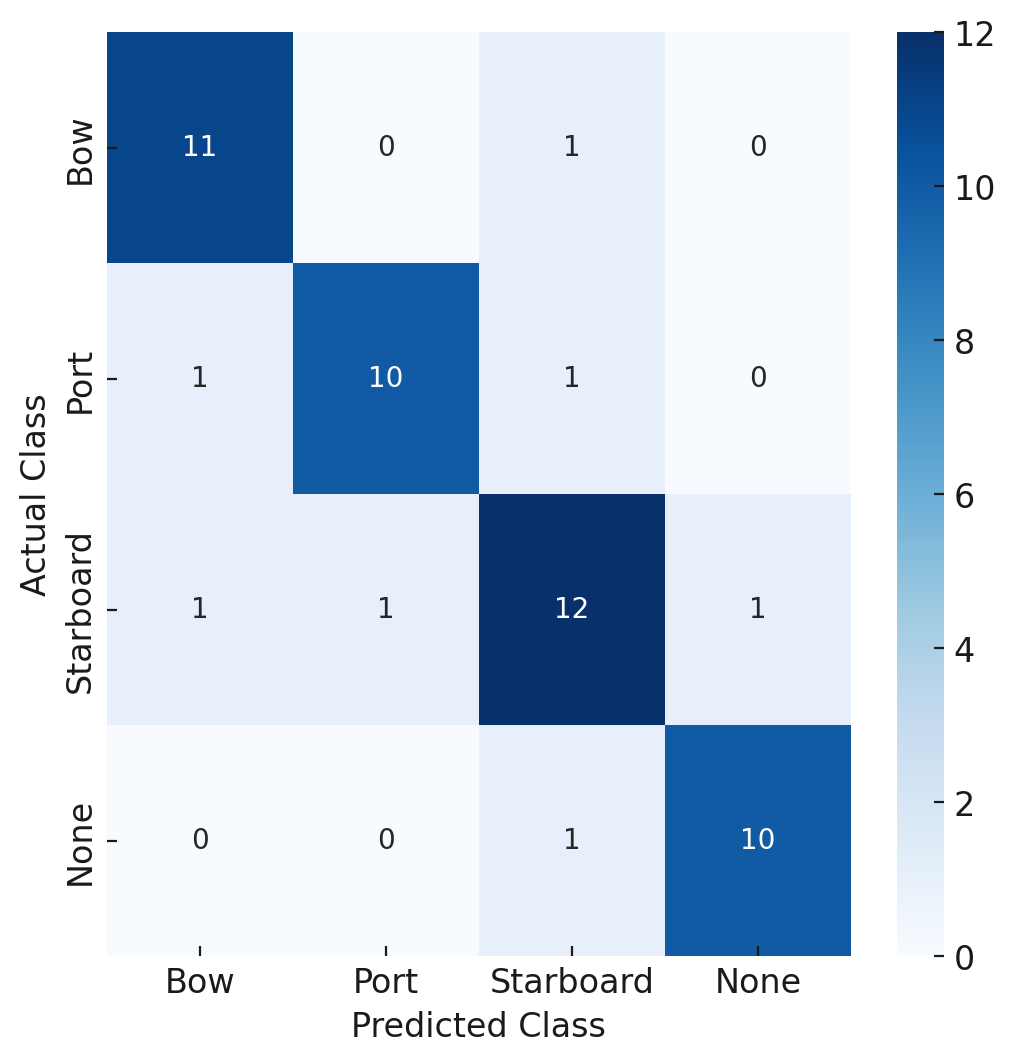}
    \includegraphics[width=0.51\columnwidth]{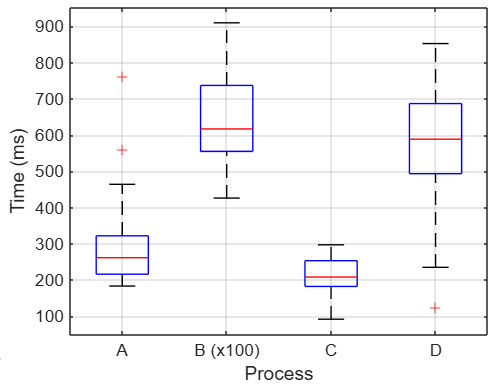}
    \caption{Confusion matrix for USV collision classification (left) and Processing time/latency analysis across layers (right).}
    \label{fig_confu}
\end{figure}

The classification accuracy of the CNN model, based on the MobileNetV2 framework, is evaluated using a confusion matrix derived from 50 trials, ensuring a balanced dataset with at least 10 trials conducted for each impact type to minimize bias and ensure fairness in the classification results. The results indicated that the proposed system successfully distinguishes between all collision types, as shown in Figure~\ref{fig_confu}.

Table~\ref{tab3} summarizes the classification performance across four distinct impact cases. Overall, the system achieved 86\% aggregate accuracy (43/50). The unified precision, recall, and F1-score reflected balanced predictive capabilities across all classes.

\begin{table}[ht]
\centering
\caption{Classification Performance Summary}
\renewcommand{\arraystretch}{1.2} 
\setlength{\tabcolsep}{3pt} 
\begin{tabular}{|c|c|c|c|c|c|c|}
\hline
\textbf{Impact Case} & \textbf{Total Cases} & \textbf{Correct} & \textbf{Precision} & \textbf{Recall} & \textbf{F1-Score} \\
\hline
Bow & 12 & 11 & 0.846 & 0.917 & 0.880\\
Port & 12 & 10 & 0.909 & 0.833 & 0.870\\
Starboard & 15 & 12 & 0.800 & 0.800 & 0.800 \\
None & 11 & 10 & 0.909 & 0.909 & 0.909\\
\hline
\textbf{Overall} & 50 & 43 & 0.861 & 0.860 & 0.859\\
\hline
\end{tabular}
\label{tab3}
\end{table}

Individually, the \lq None\rq and \lq Port\rq impacts showed the highest precision (0.909), indicating strong accuracy in correctly classifying these specific scenarios. The "Bow" impact achieved the highest recall (0.917), demonstrating reliable detection of actual collision events. However, "Starboard" impacts showed lower performance, with an F1-score of 0.800, suggesting that classification accuracy for this class could benefit from additional training or refined feature extraction.

Technically, slight variations in collision force and angle occasionally cause detection and classification errors. To address this issue, a continuous CNN model retraining facilitated by the Cloud Layer is recommended for enhancing system performance. Moreover, integrating attention mechanisms into the MobileNetV2 architecture could enhance the model's focus on critical scalogram features. For instance, H. Cheng et al. \cite{cheng2024enhanced} demonstrate that a fused spatial-channel attention mechanism in MobileNet improves classification accuracy by prioritizing relevant image regions. Moreover, sensor placement influences detection and classification outcomes, as a centrally positioned IMU ensures stable and reliable data capture, minimizing external effects. In summary, the results underscore the effectiveness and reliability of the proposed architecture while identifying areas such as the "Starboard" class that require further optimization to enhance overall system performance.

\subsection{Processing Time and System Latency Analysis}\label{sec:sec5.B}

System latency and processing times are measured across the three computational layers: Cloud (data aggregation), Edge (AI-driven classification), and IoT (data acquisition), as illustrated in Table~\ref{tab4} and Figure~\ref{fig_confu}. The results analysis across layers underscored the effectiveness of adopting Cloud-Fog Automation within the proposed architecture. The IoT Layer exhibited negligible processing latency, whereas IoT-to-Edge transmission introduced moderate latency ($<$ 300ms), reflecting the typical delays encountered in cross-regional network communication. Edge-to-Cloud latency varies due to geographical proximity and server location differences.

Notably, the Edge Layer achieved minimal processing latency (6.44 ms), ensuring reliable low-latency AI-driven collision detection and classification. Edge-to-Cloud communication latency varied substantially by cloud server location, with Australia-based cloud services exhibiting lower RTT values (210.15 ms) compared to Singapore-based services (583.66 ms) due to geographical distance and network constraints.

RTT measurements are utilized rather than one-way latency measurements due to their practicality and reliability in accurately assessing communication delays. One-way latency measurement often requires precise clock synchronization between geographically distributed nodes, introducing complexity and potential measurement inaccuracies~\cite{Shin2011}. RTT inherently accounts for both transmission and response delays, providing a more straightforward and representative measure of responsiveness and network communication performance~\cite{Lv2023}.


\begin{table}[!ht]
\centering
\small
\renewcommand{\arraystretch}{1.3}
\caption{Processing Time and System Latency Summary}
\label{tab4}
\begin{tabular}{|l|c|c|}
\hline
\textbf{Process} & \textbf{Mean (ms)} & \textbf{Std Dev. (ms)} \\ 
\hline
IoT Layer (VN) & 1.00 & ± 1.00 \\ 
\hline
A -- IoT-to-Edge (VN $\rightarrow$ AU) & 282.96 & ± 100.79 \\ 
\hline
B -- Edge Layer (AU) & 6.44 & ± 1.08 \\ 
\hline
C -- Edge-to-Cloud (AU $\leftrightarrow$ AU*) & 210.15 & ± 47.79 \\ 
\hline
D -- Edge-to-Cloud (AU $\leftrightarrow$ SG*) & 583.66 & ±  152.11 \\ 
\hline
\end{tabular}
\begin{flushleft}
\footnotesize \textit{VN: Vietnam; AU: Australia; SG: Singapore -- Locations $|$ * Cloud Service}
\end{flushleft}
\end{table}

These findings emphasize the importance of geographically optimized cloud service selection to maintain low-latency communication within USV collision classification systems. Nevertheless, the latency observed between the Edge Layer and Cloud Layer does not negatively impact low-latency collision detection and classification capabilities, as critical AI-driven processing tasks are executed efficiently at the Edge Layer. Collectively, the results validated the robustness and effectiveness of the proposed distributed Cloud-Edge-IoT computing architecture for maritime autonomous operations.

\section{Conclusions and Future Work}\label{sec:sec6}
In this paper, we proposed and developed a distributed Cloud-Edge-IoT computing architecture tailored for low-latency AI-driven collision detection and classification in USVs within the context of maritime ICPS. By leveraging the Cloud-Fog Automation paradigm, we proposed a computational framework that is distributed across decentralized hierarchical layers to effectively mitigate the limitations of the conventional onboard processing approach. 


We integrated Wavelet-based feature extraction with a MobileNetV2-based CNN model, achieving 86\% accuracy. Our experimental validation demonstrated effective computational performance, reduced onboard processing loads, and improved low-latency collision classification performance. Specifically, the IoT Layer reliably captured IMU data packages, the Edge Layer executed low-latency classification, and the Cloud Layer efficiently aggregated data for future enhancements. Experimental results analysis further confirmed the efficiency of the proposed distributed computing architecture, achieving low processing and transmission times at observed IoT and Edge layers (\textit{mean: 282.96 + 6.44 ms, SD: 100.79 + 1.08 ms}). Network latency assessments revealed variations depending on cloud server location, emphasizing the importance of geographic proximity. Specifically, RTT latency for local (AU-based) communication (\textit{mean: 210.15 ms}) is significantly lower than international cloud locations, such as Singapore (AU-SG, \textit{mean: 583.66 ms}).

Our proposed architecture not only demonstrates the potential for maritime autonomy but also provides a generalized architecture applicable to broader ICPS applications. By decentralizing computational processes and reducing latency, our framework can support diverse ICPS applications, including industrial networked robotics, precision manufacturing, industrial automation, resource management, and smart factory. For instance, industries utilizing networked robotics and autonomous systems can benefit from enhanced low-latency decision-making, advanced analytics, predictive maintenance, and dynamic system adaptation, leveraging distributed computing enabled by the proposed architecture. Moreover, critical sectors such as healthcare, infrastructure monitoring, and disaster response can utilize Cloud-Fog Automation's scalability and low-latency advantages to achieve robust teleoperation and AI-empowered autonomous capabilities~\cite{Tran2025, Nguyen2024HMI, Luu2024}.

Our work contributes to the future development of flexible, scalable, and resilient intelligent systems. The proposed architecture offers a foundational framework that can be adapted to a wide range of industrial sectors, thereby advancing intelligent scalability and operational efficiency in next-generation ICPS. Future work will extend our architecture's capabilities beyond data aggregation, hence incorporating advanced predictive analytics, federated learning, and continuous AI model fulfillment to enhance low-latency decision-making, adaptive autonomous task offloading, and proactive autonomy~\cite{Zhang2025}.


\bibliographystyle{ieeetr}
\bibliography{Bibliography}
\end{document}